\documentclass[twoside,11pt]{article}

\usepackage{blindtext}
\usepackage{enumitem}
\usepackage{color}
\usepackage{booktabs}
\usepackage{float}
\usepackage{placeins}
\usepackage{listings}
\usepackage{subcaption}
\usepackage[export]{adjustbox}
\usepackage{amsmath}
\usepackage{tikz}
\usepackage{graphicx}
\usetikzlibrary{arrows.meta, positioning}
\usepackage{jmlr2e}

\usepackage{lastpage}
\jmlrheading{23}{2026}{1-\pageref{LastPage}}{2/26; Revised ?/??}{?/??}{21-0000}{Robert D. Lee, Raymond K. W. Wong and Yang Ni}

\ShortHeadings{Bayesian Causal Discovery with Non-Gaussian Directed Acylic and Cycle Graphs}{Lee, Wong, and Ni}
\firstpageno{1}

\begin{document}

\title{\texttt{cyclinbayes}: Bayesian Causal Discovery with Linear Non-Gaussian Directed Acyclic and Cyclic Graphical Models}

\author{\name Robert D. Lee \email robderdylee@stat.tamu.edu \\
       \addr Department of Statistics\\
        Texas A\&M University\\
        College Station, TX 77843, USA.
       \AND
       \name Raymond K. W. Wong \email raywong@tamu.edu \\
       \addr Department of Statistics\\
        Texas A\&M University\\
        College Station, TX 77843, USA.
        \AND
       \name Yang Ni  \email yang.ni@austin.utexas.edu \\
       \addr Department of Statistics and Data Sciences\\
        The University of Texas at Austin\\
        Austin, TX 78705, USA.}

\editor{My editor}

\maketitle

\begin{abstract}
We introduce \texttt{cyclinbayes}, an open-source R package for discovering linear causal relationships with both acyclic and cyclic structures. The package employs scalable Bayesian approaches with spike-and-slab priors to learn directed acyclic graphs (DAGs) and directed cyclic graphs (DCGs) under non-Gaussian noise. A central feature of \texttt{cyclinbayes} is comprehensive uncertainty quantification, including posterior edge inclusion probabilities, posterior probabilities of network motifs, and posterior probabilities over entire graph structures. Our implementation addresses two limitations in existing software: (1) while methods for linear non-Gaussian DAG learning are available in R and Python, they generally lack proper uncertainty quantification, and (2) reliable implementations for linear non-Gaussian DCG remain scarce. The package implements computationally efficient hybrid MCMC algorithms that scale to large datasets. 
Beyond uncertainty quantification, we propose a new decision-theoretic approach to summarize posterior samples of graphs, yielding principled point estimates based on posterior expected loss such as posterior expected structural Hamming distance and structural intervention distance. The package, a supplementary material, and a tutorial are available on GitHub at \url{https://github.com/roblee01/cyclinbayes}.


\end{abstract}


\section{Introduction}
Causal discovery in high-dimensional settings has gained significant attention in recent years. A central line of work is based on the non-Gaussian noise assumption, as in LiNGAM and its variants \citep{shim2011lingam,shim2011direct} for directed acyclic graphs (DAGs), which are available in R packages such as \texttt{pcalg} \citep{pcalg} and \texttt{rlingam} \citep{rlingam}, and LiNG \citep{lacerda2008discovering} for directed cyclic graphs. However, these methods provide only point estimates of causal graphs, offering no mechanism for structural or parameter uncertainty quantification. Consequently, users receive limited guidance on the reliability of the inferred graph or causal effects. In particular, while many tools support non-Gaussian linear DAG discovery, software for linear non-Gaussian DCGs (feedback or reciprocal effects) is much more limited, often requiring stronger modeling restrictions (e.g., Gaussian errors or additional constraints). Moreover, publicly available software for LiNG style cyclic discovery is limited, and some earlier implementations are no longer actively maintained. While a Bayesian framework can address these limitations, summarizing the posterior over graph structures is itself challenging, as the posterior often contains many nearly equivalent graphs, making it nontrivial to identify a single representative structure in a principled way.

Therefore, we introduce \texttt{cyclinbayes}, an open-source R package for Bayesian causal discovery in high-dimensional settings. The package supports learning both DAGs and DCGs. Its key contributions are: {($i$)} an MCMC algorithm for linear non-Gaussian DAGs (Bayesian LiNGAM), {($ii$)} an MCMC algorithm for linear non-Gaussian DCGs (Bayesian LiNG),
{($iii$)} comprehensive posterior uncertainty quantification, including edge inclusion probabilities, network motif inclusion probabilities, and credible intervals for direct causal effects, and
{($iv$)} a decision-theoretic graph selection procedure based on weighted medoids under the structural Hamming distance (SHD), the structural intervention distance (SID), or any user-defined loss functions. Implemented in Rcpp, \texttt{cyclinbayes} leverages optimized C++ routines to handle large-scale, high-dimensional datasets with substantial computational efficiency.

\section{Model Specification}
Consider $p$ random variables and $n$ observations. Let $Y_{i}^{(q)}$ denote the $i$th variable of observation $q$. We model them with a linear structural equation model (SEM) with non-Gaussian noise,
\begin{align}\label{eq:sem}
  \textstyle Y^{(q)}_{i} = \sum_{j \in \mathrm{pa}(i)} B_{ij} Y^{(q)}_{j} + \epsilon^{(q)}_{i},
\end{align}
where $\mathrm{pa}(i)$ denotes the parent set of node $i$ in a graph $G$, the entries of $B = (B_{ij})_{i,j=1}^{p}$ are the direct causal effects, and $\epsilon_i^{(q)}$ is a noise variable drawn from a finite mixture of Gaussians.  
If the underlying graph $G$   is acyclic, the SEM is recursive.


To facilitate sparse structure learning and interpretable posterior inference, we place hierarchical priors on both the graph adjacency indicators and the causal effect coefficients. Edge inclusion is governed by a Beta-Bernoulli prior on adjacency indicators $E_{ij}$ (where $E_{ij}=1$ if $j\to i$),
\[E_{ij} \mid \gamma \sim \mathrm{Bernoulli}(\gamma),~~ \gamma \sim \mathrm{Beta}(a_\gamma, b_\gamma),\]
where $\gamma$ is the prior probability of edge inclusion. Conditional on $E_{ij}$, the causal effect $B_{ij}$ follows a spike and slab prior, with a point mass at zero for excluded edges and a Gaussian slab with variance $\gamma_1$ (itself given an inverse gamma prior) for included edges,
\[B_{ij}\mid E_{ij},\gamma_1 \sim (1 - E_{ij})\,\delta_0 + E_{ij}\,\mathcal{N}(0,\gamma_1),~~ \gamma_1 \sim \mathrm{InverseGamma}(a_{\gamma_1}, b_{\gamma_1}).\]
This prior is conjugate when the graph is acyclic. 
\section{Package Implementation}
Our package \texttt{cyclinbayes} provides two fast Bayesian samplers for DAGs and DCGs based on the model in Section 2, along with tools for posterior uncertainty quantification and analysis. Both samplers use MCMC to update the graph structure and causal coefficients, and are implemented in Rcpp for speed. Figure 1 outlines the workflow: run a sampler, then compute a graph point estimate, credible intervals for causal effects, and posterior probabilities for user-specified network motifs. The major functions are:

\begin{enumerate}
    \item \texttt{BayesDAG()} implements Bayesian LiNGAM for learning acyclic causal structures by restricting the graph in \eqref{eq:sem} to be a DAG. The algorithm uses a collapsed Gibbs sampler in which the causal effect coefficients are marginalized out to improve mixing, together with simulated annealing to mitigate local optima. Given a data matrix and prior hyperparameters, the function returns posterior samples of the adjacency matrix and all associated model parameters.
    \item \texttt{BayesDCG()} implements Bayesian LiNG for DCG strucutre learning using a Gibbs-within-Metroplis algorithm. Input and output are similar to \texttt{BayesDAG()}. 
     \item \texttt{point\_est\_graph()} computes a decision-theoretic posterior point estimate of the adjacency matrix from posterior samples under a chosen distance metric (SHD, SID, or a user-specified distance). SID is only applicable to DAGs.
    \item \texttt{posterior\_interval\_est()}
    computes the Highest Posterior Density (HPD) and equal-tailed credible intervals (at a user-chosen level) for the model parameters such as direct causal effects.
   
    \item \texttt{posterior\_network\_motif()} computes the posterior probability mass of a user-specified network motif (i.e., a subgraph of a specific set of nodes) by checking how often all its edges appear simultaneously in the posterior graph samples.
\end{enumerate}

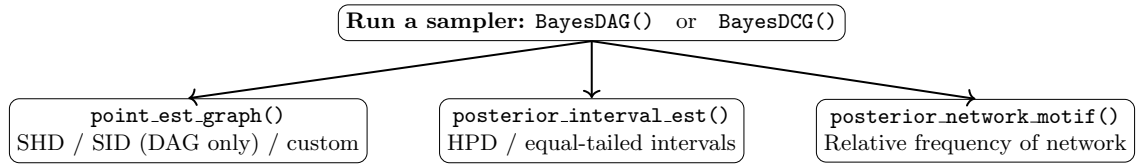
\begin{figure}[H]\label{flow}
\centering
\scalebox{0.95}{
\begin{tikzpicture}[
  node distance = 8mm and 10mm,
  box/.style = {draw, rounded corners, align=center, inner sep=3pt, font=\footnotesize},
  arrow/.style = {->, thick}
]

\node[box] (sampler)
{\textbf{Run a sampler:} \texttt{BayesDAG()} \; or \; \texttt{BayesDCG()}};

\node[box, below=of sampler] (f2) {\texttt{posterior\_interval\_est()}\\
\footnotesize HPD / equal-tailed intervals};
\node[box, left=of f2] (f1)
{\texttt{point\_est\_graph()}\\
\footnotesize SHD / SID (DAG only) / custom};
\node[box, right=of f2] (f3)
{\texttt{posterior\_network\_motif()}\\
\footnotesize Relative frequency of network};

\draw[arrow] (sampler.south) -- (f1.north);
\draw[arrow] (sampler.south) -- (f2.north);
\draw[arrow] (sampler.south) -- (f3.north);

\end{tikzpicture}
}
\caption{Typical \texttt{cyclinbayes} workflow: run a sampler, then perform graph selection, interval estimation, and motif posterior analysis.}
\end{figure}

\paragraph{Decision-Theoretical Approach for Graph Selection}
Selecting a single graph from a Bayesian posterior is challenging because the posterior over graph structures is often highly multimodal. Standard summaries, such as the Maximum A Posteriori (MAP) graph or edge-wise thresholding, either focus on a single mode or ignore global structural constraints (e.g., the resulting graph may not even be DAGs). We therefore adopt a Bayesian decision-theoretic framework in which the graph point estimator  minimizes posterior expected loss under a chosen graph discrepancy:
\begin{align}\label{eq:optim}
    &a^{*} = \arg\min_{a}E[d(a,{\cal{G}})|\text{data}] \approx \arg\min_{a} \frac{1}{m}\sum_{h=1}^{m}d(a,{\cal{G}}^{(h)}),
\end{align}
where the expectation is the posterior expectation with respect to $\cal{G}$, which is approximated by Monte Carlo via the posterior samples$\{{\cal{G}}^{(1)},\ldots,{\cal{G}}^{(m)}\}$. Because the space of possible graphs is vast, we approximate the optimization problem in \eqref{eq:optim} using the same set of posterior samples of graphs. This naturally leads to the posterior weighted medoid, which selects the graph minimizing expected loss among the sampled structures.
More specifically, let $\{{\cal{G}_{*}}^{(1)},\ldots,{\cal{G}_{*}}^{(v)}\}$ denote the set of unique graph structures among the $m$ posterior samples. Each unique graph ${\cal{G}_{*}}^{(u)}$ is assigned a weight $w_{u}$, $u=1,\ldots,v$, 
equal to its posterior probability. For each candidate graph ${\cal{G}_{*}}^{(l)}$, we compute its total weighted distance $D_{l}$ to all other unique graphs:
\begin{align*}
    &D_{l} = \sum_{u=1}^{v}w_{u}d({\cal{G}_{*}}^{(l)},{\cal{G}_{*}}^{(u)}),
\end{align*}
where \( d(\cdot,\cdot) \) measures the discrepancy between the two graphs. The weighted medoid is then ${\cal G}_{*}^{(k)}$ where $k = \arg\min_{l \in \{1,\dots,v\}} D_l$, which is an approximate solution to \eqref{eq:optim}.
We provide multiple options for distance metrics \( d(\cdot,\cdot) \) including SHD, SID (for DAGs only), and any user-specified distance. 



By operating on the posterior distribution of the graphs rather than the marginal posterior inclusion probability of each individual edge, this new joint decision rule fully leverages posterior structural uncertainty and yields a point estimate that reflects the overall posterior consensus.

\section{Conclusion}

We have introduced \texttt{cyclinbayes}, an R package that provides a unified Bayesian framework for causal discovery under linear non-Gaussian structural equation models. For DAGs, our package offers a scalable, fully Bayesian treatment of the LiNGAM model, enabling comprehensive uncertainty quantification of both direct causal effects and user-specified network motifs, which are capabilities absent from existing implementations. \texttt{cyclinbayes} also represents one of the few available software implementations for linear non-Gaussian DCG learning and, to our knowledge, the only Bayesian implementation, thereby uniquely enabling uncertainty quantification in settings with feedback loops.

A key methodological contribution is the novel decision-theoretic approach to posterior graph summarization. Rather than relying on edge-wise thresholding or MAP estimation, which can yield graphs that poorly represent the posterior distribution, our approach selects a representative graph that minimizes posterior expected loss under structural discrepancy metrics such as SHD and SID. This principled approach fully leverages the rich structural uncertainty captured by Bayesian inference and produces point estimates that reflect global posterior consensus.

The computationally efficient Rcpp implementation, combined with comprehensive tools for posterior inference and graph selection, makes \texttt{cyclinbayes} a practical resource for applied researchers seeking flexible causal modeling with rigorous uncertainty quantification.

\vskip 0.2in
\bibliography{reference}

\end{document}